\documentclass{svproc}
\usepackage{url}

\usepackage{amsmath}
\usepackage{amssymb}
\usepackage{graphicx}
\usepackage{csquotes}
\usepackage{multirow}
\usepackage{tabularx}
\usepackage{xcolor} 
\usepackage[colorlinks]{hyperref} 
\usepackage{mathtools}
\usepackage{wasysym}

\newcommand{\Mod}[1]{\ \mathrm{mod}\ #1}

\usepackage{amsmath}
\DeclareMathOperator*{\argmax}{arg\,max}

\definecolor{green}{HTML}{48bf91}
\definecolor{gray}{HTML}{605e5e}

\begin{document}
\mainmatter              
\title{Collaborative search and autonomous task allocation in organizations of learning agents}
\titlerunning{Collaborative search and autonomous task allocation}  
%
\author{Stephan Leitner}
\authorrunning{Stephan Leitner} 
%
\tocauthor{Stephan Leitner}
\institute{University of Klagenfurt, Klagenfurt, Austria,\\
\email{stephan.leitner@aau.at}}

\maketitle              

\begin{abstract}

This paper introduces a model of multi-unit organizations with either static structures, i.e., they are designed top-down following classical approaches to organizational design, or dynamic structures, i.e., the structures emerge over time from micro-level decisions. In the latter case, the units are capable of learning about the technical interdependencies of the task they face, and they use their knowledge by adapting the task allocation from time to time. In both static and dynamic organizations, searching for actions to increase the performance can either be carried out individually or collaboratively. The results indicate that \textit{(i)} collaborative search processes can help overcome the adverse effects of inefficient task allocations as long as there is an internal fit with other organizational design elements, and \textit{(ii)} for dynamic organizations, the emergent task allocation does not necessarily mirror the technical interdependencies of the task the organizations face, even though the same (or even higher) performances are achieved.

\end{abstract}
\section{Introduction}

Designing organizations includes a multiplicity of decisions, such as breaking down the task of the larger problem for smaller units, allocating responsibility and authority to departments and individuals, coordinating behavior through incentives, communication, leadership, and routines, among others, and it is well known that an organization's design substantially impacts the organization's performance \cite{burton2018,burton2020}. The main challenges of organizational design are to achieve an external fit, i.e., to design organizations for dynamic and uncertain situations and perhaps even situations that have not been seen before \cite{burton2018}, and an internal fit among the organizational design elements \cite{thompson2017}, which might be particularly difficult when organizations evolve through phases of their life-cycle and the employees' capabilities and knowledge are dynamic \cite{cardinal2004}.

There are two main world-views on organizational design: First, classic approaches follow the premise of the rational actor and postulate that organizational design is the result of \textit{deliberate} decisions \cite{tsoukas1993}; following this view, managers design feasible organizations top-down. Second, evolutionary approaches consider that organizational structures emerge bottom-up. The latter approach includes a shift from the macro-level to the micro-structures, focusing on mechanisms that drive the emergence of organizational design elements \cite{joseph2018}. This paper addresses two such micro-level issues: First, limited information, learning, and adaptation, and second, collaborative search processes.

\textit{Limited information, learning, and adaptation} concern the technical characteristics and decomposition of the task the organization faces.
Previous research recommends that an organization's structure should mirror the task's technical interdependencies (mirroring hypothesis) \cite{sanchez1996}. There are ambiguous results regarding this hypothesis; some previous research criticizes it based on empirical evidence, and, at the same time, there are also empirical results that support it \cite{baldwin2014,querbes2018}. Efficiently designing organizations top-down and in line with the mirroring hypothesis requires that the technical structure (i.e., the structure of interdependencies) is public knowledge. In reality, this structure is unknown and unclear in most cases \cite{raveendran2020}. Highly complex tasks might not only be challenging to decompose; previous research argues that increasing the number of interdependencies also unfolds non-linear effects that lead to performance drops, what is often labelled as \enquote*{complexity catastrophe} \cite{kauffman1993}. This paper addresses both cases of organizational design mentioned before; there are scenarios in which \textit{(i)} the technical interdependencies of the task are known beforehand, and organizations are designed top-down, and \textit{(ii)} the technical interdependencies are not known, but agents learn about it over time and can adapt the task allocation over time. 

This paper relies on situated learning theory to model \textit{collaborative search processes}, according to which search processes might take place in interactive communities \cite{yuan2004}. While traditional search algorithms mainly focus on individual search processes \cite{wall2020}, this paper enriches the models of an organization with distributed and autonomous decision-makers by a social network that constitutes organizational connections. These connections are then used to autonomously coordinate search behavior, resulting in collaborative search efforts. For dynamic and static organizations, the paper tests whether there are organizational design elements, such as control mechanisms and (collaborative) search processes, that either reinforce or weaken the \enquote*{complexity catastrophe}. 

The remainder of this paper is organized as follows: Sec. \ref{sec:model} introduces the model and the method of data analysis, Sec. \ref{sec:results} presents and discusses the results. Finally, Sec. \ref{sec:conclusion} summarizes and concludes the paper. 

\section{Model}
\label{sec:model}
 
The model builds on the well-known $N\!K$ framework \cite{wall2020}. The organization comprises $M\in \mathbb{N}$ organizational units, referred to as agents henceforth. All agents face an $N$-dimensional decision problem with $K$ interdependencies among them, where $N\in \mathbb{N}$ and $K\in \mathbb{N}_0$. The interdependencies shape the decision problem's complexity. Due to limited capacities, the agents cannot solve the entire decision problem alone, but they decompose it into $M$ sub-problems that agents can handle (Sec. \ref{sec:task-environment}). The agents aim to increase their utilities by employing an individual or collaborative search processes (Sec. \ref{sec:decisions}). The agents know that they face a complex decision problem. However, they do not know the actual number and structure of interdependencies between decisions. Still, they are endowed with the capability to learn about the structure of interdependencies (Sec. \ref{sec:learning}). Also, the agents use their knowledge by adapting the task allocation from time to time (Sec. \ref{sec:auction}). For $t=\{1,\dots,T\}\subset \mathbb{N}$ periods it is observed how the agents' decisions affect the organization's performance. 
The model was implemented in Matlab\textsuperscript{\textregistered} (R2022a). 

\subsection{Task environment and decomposition}
\label{sec:task-environment}
The decision problem faced by the agents consists of $N$ binary decisions and is formalized by 
$\mathbf{d}=\left[d_1, d_2, \dots, d_N \right]$,
where $d_n\in \{0,1\}$ and $n=\{1,\dots,N \} \subset \mathbb{N}$. Every decision $d_n$ contributes $f(d_n) \sim U(0,1)$ to the organization's performance. Due to interdependencies among decision, the performance contribution $f(d_n)$ might not only be affected by decision $d_n$ but also by $K$ other decisions. The corresponding contribution function for decision $d_n$ is formalized by 
    $f\left(d_n\right)=f\left(d_n, d_{i_1}, \dots, d_{i_K} \right)$,
where $\{i_1, \dots, i_K\} \subseteq \{1, \dots, n-1, n+1,\dots,N\}$ and $0\leq K \leq N-1$. The organizations' performance is the average of all performance contributions:
\begin{equation}
    \label{eq:performance}
    P(\mathbf{d})=\frac{1}{|\mathbf{d}|} \sum_{n=1}^{|\mathbf{d}| }f\left(d_n\right)~.
\end{equation}

The agents are limited in their capabilities and/or resources, i.e., they might have limited cognitive capacities, limited time, or limited further resources to solve the decision problem. Consequently, they have to collaborate to find a feasible solution to the complex decision problem captured by the task environment. To do so, they decompose the decision problem into $M$ sub-problems $\mathbf{d_m}$, where $m=\{1,\dots,M\}\subset\mathbb{N}$ and $[\mathbf{d}_1,\dots,\mathbf{d}_M]=\mathbf{d}$. For agent $m$, the decisions $\mathbf{d}_m$ represent the area of responsibility, while the complement $\mathbf{d}_{-m} = \mathbf{d} \setminus \mathbf{d}_m$ is referred to as residual decisions. The agents can observe the solutions to their sub-problem $\mathbf{d}_m$ at any time. However, the solutions to the residual decision problem $\mathbf{d}_{-m}$, can only be observed \textit{after} implementation. 

\begin{figure}
\centering
\vspace{-0.2cm}
\includegraphics[width=0.8\textwidth]{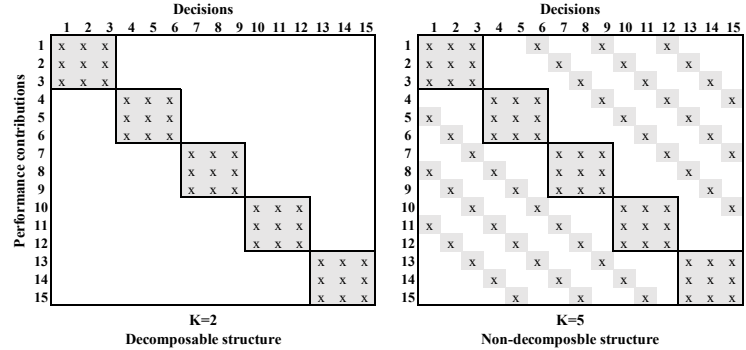}
    \caption{Interdependence matrices}
    \label{fig:matrices}
\end{figure}

This paper considers two stylized interdependence structures presented in Fig. \ref{fig:matrices}, where an \enquote*{x} indicates that a decision and a performance contribution are interdependent. The task allocation indicated by black lines is used for scenarios with \textit{top-down designed organizations}. The considered structures are of complexity $K=2$ and $K=5$, representing a fully decomposable and non-decomposable decision problem, respectively. 

In \textit{organizations with emergent structures}, the allocation of tasks to agents might be adapted from time to time, i.e., agents might swap tasks. In these scenarios, in every period $t\Mod \tau = 0$, agents can adjust the task allocation (Sec. \ref{sec:auction}). In contrast, in periods $t \Mod \tau \neq 0$, agents seek to maximize their utility given the currently active task allocation (Sec. \ref{sec:decisions}), where $\tau\in \mathbb{N}$. The task allocation in period $t=1$ follows a random process that allocates tasks equally so that the number of decisions assigned to agent $m$ is $|\mathbf{d}_m|=N/M$. 
\subsection{Utility functions and search processes}
\label{sec:decisions}
The performance contributions of agent $m$'s own ($\mathbf{d}_{mt}$) and residual decisions ($\mathbf{d}_{-mt}$) in $t$ are denoted by $P(\mathbf{d}_{mt})$ and $P(\mathbf{d}_{-mt})$, respectively.
The organization employs a linear outcome-based incentive scheme that shapes the agents' utility functions. In particular, the parameter $\alpha\in \mathbb{R^+}$ is used to weight the agents' own and residual performances, respectively, where $0 \leq \alpha \leq 1$. Agent $m$'s utility at period $t$ is formalized by 
\begin{equation}
\label{eq:utility}
    U_{}(\mathbf{d}_{mt},\mathbf{d}_{-mt}) = \alpha \cdot P\left( \mathbf{d}_{mt} \right) + \left(1-\alpha\right) \cdot  P\left( \mathbf{d}_{-mt} \right) ~.
\end{equation}
The agents seek to maximize their utilities by employing one of the following two variants of a hill-climbing algorithm: 

\subsubsection{Individual search.}

Agent $m$ discovers a solution $\mathbf{d}^{\ast}_{mt}$ to their partial decision problem in period $t$ characterized by a Hamming distance of $1$ to the solution $\mathbf{d}_{mt-1}$, i.e., $\mathbf{d}_{mt}^{\ast}$ is different from $\mathbf{d}_{mt-1}$ in exactly one position.
Direct communication between agents is omitted in individual hill-climbing, so agent $m$ has no information about the other agents' decisions but relies on the other agents' decisions from the previous period, $\mathbf{d}_{-mt-1}$, to compute the utility. Agent $m$ selects the solution to be implemented in $t$ from their options $\mathbf{D}^{ind}_t = \{\mathbf{d}_{mt-1},\mathbf{d}^{\ast}_{mt} \}$ according to the following rule:

\begin{equation}
\label{eq:individualwalk}
    \mathbf{d}_{mt} = \argmax_{\mathbf{d}^{'} \in \mathbf{D}^{ind}_t} ~U\!\left(\mathbf{d}^{'}, \mathbf{d}_{-mt-1} \right)~. 
\end{equation}

\subsubsection{Collaborative search.} Agents are connected in a ring network, and they interact with one of their nearest neighbors with probability $\mathbb{P}$. If they interact, agents $m$ and $n$ jointly perform adjacent hill-climbing \cite{yuan2004} to maximize their \textit{joint utility}. They share information about the solutions $\mathbf{d}_{m}$ and $\mathbf{d}_{n}$ to their partial decision problem. Let us denote the solutions to the decisions outside the two agents' areas of responsibility by $\mathbf{d}_{-(m,n)} = \mathbf{d} \setminus (\mathbf{d}_{m} \cup \mathbf{d}_{n})$. Then, the agents' joint utility in period $t$ is the mean of the individual utilities in Eq. \ref{eq:utility}:

\begin{equation}
    U^{adj} 
    \left( 
    \mathbf{d}_{mt}, 
    \mathbf{d}_{nt}, 
    \mathbf{d}_{-(m,n)t} 
    \right) 
    =  
    \frac{1}{2} \cdot 
    ( 
    U ( 
    \mathbf{d}_{mt}, 
    \underbrace{\mathbf{d}_{-mt}}_{\mathclap{\mathbf{d}_{-(m,n)t} \cup \mathbf{d}_{nt}}}
    )
   +
   U ( 
   \mathbf{d}_{nt}, 
   \underbrace{\mathbf{d}_{-nt}}_{\mathclap{\mathbf{d}_{-(m,n)t} \cup \mathbf{d}_{mt}}} 
   ) 
   )
\end{equation}

\noindent The two agents discover and share with their counterparts the new solutions $\mathbf{d}^{\ast}_{mt}$ and $\mathbf{d}^{\ast}_{nt}$. Again, the newly discovered solutions are characterized by a Hamming distance of $1$ to the corresponding solutions in the previous period. For the decisions outside their areas of responsibility, the agents $m$ and $n$ rely on the residual solutions implemented in the last period, $\mathbf{d}_{-(m,n)t-1}$. The agents jointly choose the solutions to be implemented in period $t$ from the tuples $\mathbf{D}^{adj}_t = \{ (\mathbf{d}_{mt-1}, \mathbf{d}_{nt-1}), (\mathbf{d}^{\ast}_{mt}, \mathbf{d}_{nt-1}), (\mathbf{d}_{mt-1}, \mathbf{d}^{\ast}_{nt})\}$ according to the rule

\begin{equation}
\label{eq:adjwalk}
    (\mathbf{d}_{mt}, \mathbf{d}_{nt}) = \argmax_{(\mathbf{d}^{'}_m, \mathbf{d}^{'}_n) \in \mathbf{D}^{adj}_t}  U^{adj}\left(\mathbf{d}^{'}_m,\mathbf{d}^{'}_n, \mathbf{d}_{-(m,n)t-1}\right)~.
\end{equation}

\subsubsection{Computation of the overall solution.}
The solution to the decision problem that is implemented in period $t$ is the concatenation of the decisions made by all $M$ agents, $\mathbf{d}_t = \left[ {\mathbf{d}_{1t}}, \dots, \mathbf{d}_{Mt}\right]$, and the performance achieved by the organization in $t$ is $P(\mathbf{d}_t)$ (Eq. \ref{eq:performance}).

\subsection{Learning mechanism} 
\label{sec:learning}

The agents know that they face a complex decision problem, but they do not know the exact structure of interdependencies among decisions. However, agents are endowed with beliefs on the interdependencies, and they update them in all periods $t \Mod \tau \neq 0$. We formalize agent $m$'s belief on the interdependencies between decisions $i$ and $j$ in period $t$ by $b_{mt}^{ij} \in \mathbb{R}$, where $i,j = \{1,\dots,N\} \subset \mathbb{N}$, $i\neq j$, and $0\leq b_{mt}^{ij} \leq 1$. The beliefs $b_{mt}^{ij}$ are computed as the mean of the Beta distribution $B( p_{mt}^{ij}, q_{mt}^{ij})$. 
For the initial beliefs, $p_{m1}^{ij} = q_{m1}^{ij} = 1$ so that $b_{m1}^{ij} =0.5$. During the observation period, agent $m$ makes decisions in their area of responsibility and fixes the decisions $\mathbf{d}_{mt}$ to be implemented in $t$ by either following the individual (Eq. \ref{eq:individualwalk}) or adjacent hill-climbing algorithm (Eq. \ref{eq:adjwalk}). If agent $m$ decides to change a decision so that $\mathbf{d}_{mt}:=\mathbf{d}_{mt}^{\ast}$, the beliefs on interdependencies are updated as follows:
\begin{enumerate}
    \item Let us denote the decision that has been flipped by agent $m$ in $t$ by $i$, where $d_{it} \in \mathbf{d_{mt}}$. After implementing the decisions $\mathbf{d}_{mt}$, agent $m$ observes the performance contributions of all decisions within their area of responsibility.
    \item Whenever agent $m$ observes that the performance contribution of decision $j$ changes from period $t-1$ to period $t$ if the decision $i$ is flipped, $p_{mt}^{ij}$ is increased by $1$, otherwise  $q_{mt}^{ij}$ is increased by $1$:  
\end{enumerate}

\begin{equation}
      \label{eq:update}
    \left(p_{mt}^{ij},q_{mt}^{ij}\right)   = 
    \begin{cases}
      \left(p_{mt-1}^{ij}+1, q_{mt-1}^{ij}\right) &   \text{if } f(d_{jt}) \neq f(d_{jt-1})~,\\[8pt]
      \left(p_{mt-1}^{ij}, q_{mt-1}^{ij}+1\right)        &              \text{otherwise .}
    \end{cases} \\
\end{equation}

\begin{enumerate}
    \setcounter{enumi}{2}   
    \item Agent $m$ recomputes the beliefs $b_{mt}^{ij}$.
\end{enumerate}
Please note that agents can only observe the performance contributions \textit{within their} areas of responsibility. Suppose the decision problem is decomposed so that there are interdependencies with decisions from \textit{outside} an agent's area of responsibility; in that case, there might be external influence on performance contributions that the agent cannot identify as such. 
\subsection{Task re-allocation mechanism}
\label{sec:auction}
In all periods $t \mod \tau = 0$, agents are granted the possibility to re-organize the task allocation.\footnote{Please note that a re-allocation of decision also affects the computation of the agent's utility in terms of what is regarded as own and residual performance (see Eq. \ref{eq:utility}).} 
To account for limitations in resources, every agent is characterized by a maximum capacity $C_{m}$ that indicates the maximum number of decisions that agent $m$ can handle at a time. $C_m$ can be interpreted in terms of maximum cognitive capacity or maximum financial resources, time, manpower, etc., that are available to solve decision problems. 

\subsubsection{Computation and exchange of signals.}
Agents follow the idea of the mirroring hypothesis and aim at maximizing the interdependencies within their own areas of responsibility. The process is organized as follows:
\begin{enumerate}
    \item Agent $m$ identifies the task $i$ in their own area of responsibility that is associated with the minimum belief on internal interdependencies. Agent $m$ also sends a signal (Eq. \ref{eq:mink}) that is used as a threshold for trading this decision, i.e., the task is only re-allocated if the other agents' signals exceed the threshold signal.
\end{enumerate}
\begin{equation}
\label{eq:mink}
  {\rho}_{mt}^{i} = \min_{\forall i: d_{it} \in \mathbf{d}_{mt}}\left( \frac{1}{|\mathbf{d}_{mt}|-1} \sum_{\substack{\forall j: d_{jt} \in \mathbf{d_{mt}} \\ j \neq i}} b_{mt}^{ij} \right)
\end{equation}
\begin{enumerate}
    \setcounter{enumi}{1}   
    \item Agent $m$ informs the other agents that the task $i$ that fulfils Eq. \ref{eq:mink} and the threshold signal ${\rho}_{mt}^{i}$. Agents $r$ proceed with the next step and send signals iff $|\mathbf{d}_{rt}| < C_r$.
    \item Agents $r$ submit the average belief on the interdependencies between the offered task $i$ with the decisions within his or her area of responsibility $\mathbf{d}_{rt}$ as a signal in period $t$. Agent $r$'s signal for decision $i$ in $t$ is formalized by
    \begin{equation}
        \label{eq:k-basedbids}
        \bar{\rho}_{rt}^{i}= \frac{1}{|\mathbf{d}_{rt}|} \sum_{{\forall j: d_{jt} \in \mathbf{d_{rt}}}} b_{rt}^{ij}
    \end{equation}
\end{enumerate}
\subsubsection{Task re-allocation.}
Once all agents sent their signals, for every offer $i$, there are at most $M-1$ signals. Recall, agent $m$ offered task $i$ at a threshold signal of ${\rho}^{i}_{mt}$ and the other agents sent signals $\bar{\rho}^{i}_{rt}$. 
Let us denote the set of signals for task $i$ in period $t$ by $P_t^i$, the maximum signal for task $i$ in period $t$ by $\bar{\rho}^{i}_{r^{\ast}t}= \max_{\bar{\rho}^{i}_{rt}\in P_t^i} (\bar{\rho}^{i}_{rt})$, and the agent sending the maximum signal by $r^{\ast}$. The tasks are (re-)allocated as follows:
If the maximum signal $\bar{\rho}^{i}_{r^{\ast}t}$ is equal to or exceeds the threshold signal ${\rho}^{i}_{mt}$, the task $i$ is re-allocated from agent $m$ to agent $r^{\ast}$ according to
    \begin{subequations}
    \begin{eqnarray}
      \mathbf{d}_{mt} & := & \mathbf{d}_{mt-1} \setminus \{ d_{it-1} \} ~\text{and}\\
        \mathbf{d}_{r^{\ast}t} & := & \left[ {\mathbf{d}_{r^{\ast}t-1}},d_{it-1}  \right]~,
        \end{eqnarray}
    \end{subequations}
    \noindent where $\setminus$ indicates the complement. 
If the maximum signal $\bar{\rho}^{i}_{r^{\ast}t}$ does not exceed the threshold ${\rho}^{i}_{mt}$, agent $m$ remains responsible for task $i$, so that 
        $\mathbf{d}_{mt}:=\mathbf{d}_{mt-1}$.

\subsection{Parameters and data analysis}

\subsubsection{Parameters.} The main parameters are summarized in Tab. \ref{tab:variables}. This paper puts particular emphasis on the analysis of the relation between task performance (as the dependent variable) and task complexity $K$, collaborative search probability $\mathbb{P}$, and the incentive parameter $\alpha$ (the independent variables). To assure comparability across simulation runs, the observed performance $P(\mathbf{d}_{ts})$ is normalized by the maximum attainable performance in that scenario, $P(\mathbf{d}^{\ast}_s)$, so that $\tilde{P}(\mathbf{d}_{ts})=P(\mathbf{d}_{ts})/P(\mathbf{d}^{\ast}_s)$.
In addition to cases in which agents can adapt the task allocation in every $\tau=25$ periods, i.e., \textit{emergent organizational structures}, there are benchmark scenarios in which the initial allocation of tasks already follows the mirroring hypothesis (which is indicated the bold lines in Fig. \ref{fig:matrices}) and the agents cannot re-allocate tasks ($\tau=\emptyset$), i.e., \textit{top-down designed organizations}. 

\begin{table}
\caption{Parameters}
\label{tab:variables}
\renewcommand{\arraystretch}{1.2}
\begin{tabular}{llll}
\\ \hline
Type                                  & Variables              & Notation                    & Values                         \\ \hline
\multirow{4}{*}{Independent variables} & Task complexity        & $K$                         & \{3, 5\}                       \\
                                   & Time steps             & $t$                         & $\{1:1:150\}$   \\
                                      & Collaborative search probability  & $\mathbb{P}$                & $\{0:0.05:0.5\}$ \\
 
                                    & Incentive parameter  & $\alpha$                & $\{0,25, 0.5, 0.75\}$ \\ \hline
Dependent variable                    & Normalized task performance       & $\tilde{P}(\mathbf{d_t})$           & $[0,1]$                     \\ \hline
\multirow{4}{*}{Other parameters}       
                                      & Number of decisions    & $N$                         & 15                             \\
                                      & Agents   & $M$                         & 5                            \\
                                                                        
                                                                        & Agents' cognitive capacities   & $C_m$                         & 5                            \\
                                                                                                             & Task re-allocation     & $\tau$                      & \{$\emptyset, 25$\}         \\
                                      & Number of simulations  & $S$                         & 800                        \\
\hline
\end{tabular}%
\end{table}

\vspace{-0.5cm}

\subsubsection{Regressions and partial dependencies.} To analyze the functional dependencies between the dependent and the independent variables, regression neural networks are trained, and partial dependencies are computed \cite{patel2018,blanco2022}.
Let $\mathbf{X}$ be the set of all independent variables included in Tab. \ref{tab:variables}. The subset $\mathbf{X}^s$ includes the independent variable(s) that are in the scope of the analysis, and $\mathbf{X}^c$ consists of the complementary set of $\mathbf{X}^s$ in $\mathbf{X}$. Then, $f(\mathbf{X})=f(\mathbf{X}^s,\mathbf{X}^c)$ represents the trained regression model. The partial dependence of the performance on the independent variables in scope is defined by the expectation of the performance concerning the complementary independent variables so that 

\begin{equation}
\label{eq:dependencies}
    f^s(\mathbf{X}^s)= E_c(f(\mathbf{X}^s,\mathbf{X}^c)) \approx \frac{1}{V}\sum_{i=1}^{V} f(\mathbf{X}^s,\mathbf{X}_{(i)}^c)~,
\end{equation}

 \noindent where $V$ is the number of independent variables in $\mathbf{X}^c$ and $\mathbf{X}_{(i)}^c$ is the $i^{th}$ element. By marginalising over the independent variables in $\mathbf{X}^c$, we get a function that depends only on the independent variables in $\mathbf{X}^s$.
\vspace{-0.5cm}
\subsubsection{Task allocation efficiency.} The efficiency of task re-allocation is evaluated using the following metric: Let $C(\mathbf{d}_{mt})$ be a count-function that returns the number of interdependencies \textit{within} agent $m$'s sub-problem in $t$. Then, the following ratio of interdependencies within agent $m$'s sub-problem (nominator) to the total number of times the decisions assigned to agent $m$ affect performance contributions (denominator) is used as the task re-allocation efficiency metric:
\begin{equation}
\label{eq:ratio}
    \eta_{mt} = \frac{C(\mathbf{d}_{mt})}{|\mathbf{d}_{mt}|\cdot K}
\end{equation}


\section{Results}
\label{sec:results}

\subsection{Effects of complexity, time, and collaborative search on performance}

\subsubsection{Complexity.} 
\begin{figure}
\includegraphics[width=\textwidth]{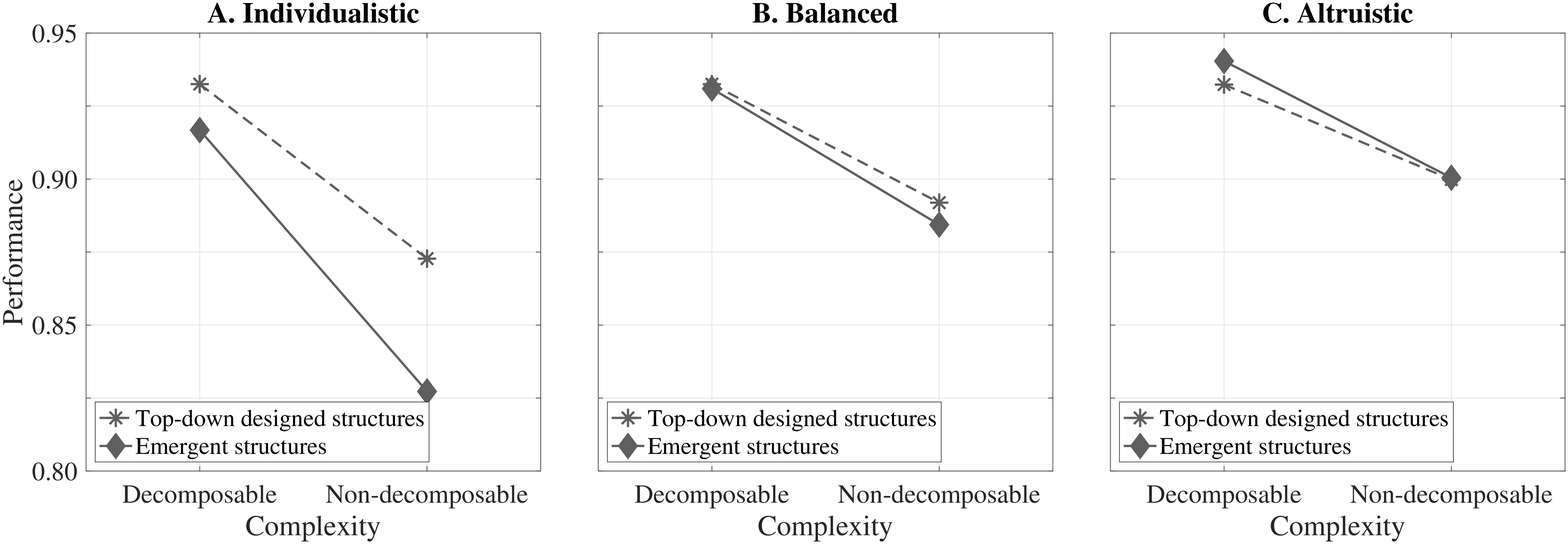}
\caption{Partial dependence of performance on complexity}
\label{fig:results-complexity}
\end{figure}

The partial dependencies of performance on complexity are plotted in Fig. \ref{fig:results-complexity}. The results indicate that whether or not endowing the agents with the capability to re-allocate tasks reinforces the \enquote*{complexity catastrophe} \cite{kauffman1993} depends on the incentive system effective in the organization. In particular, the results for top-down designed organizations reflect the finding that higher levels of complexity result in lower task performance \cite{leitner2014}. 
The results for emergent organizational structures show that individualistic incentives reinforce the effect of complexity on performance. In contrast, task re-allocation appears to slightly weaken (or, at least, not reinforce) this effect in cases with altruistic incentives. Thus, focusing on complexity only, bottom-up designed organizations are best off if they employ altruistic incentives, whereas individualistic incentives result in the most significant drop in performance.

\vspace{-0.5cm}
\subsubsection{Time and collaborative search probability.}

\begin{figure}[ht]
\includegraphics[width=\textwidth]{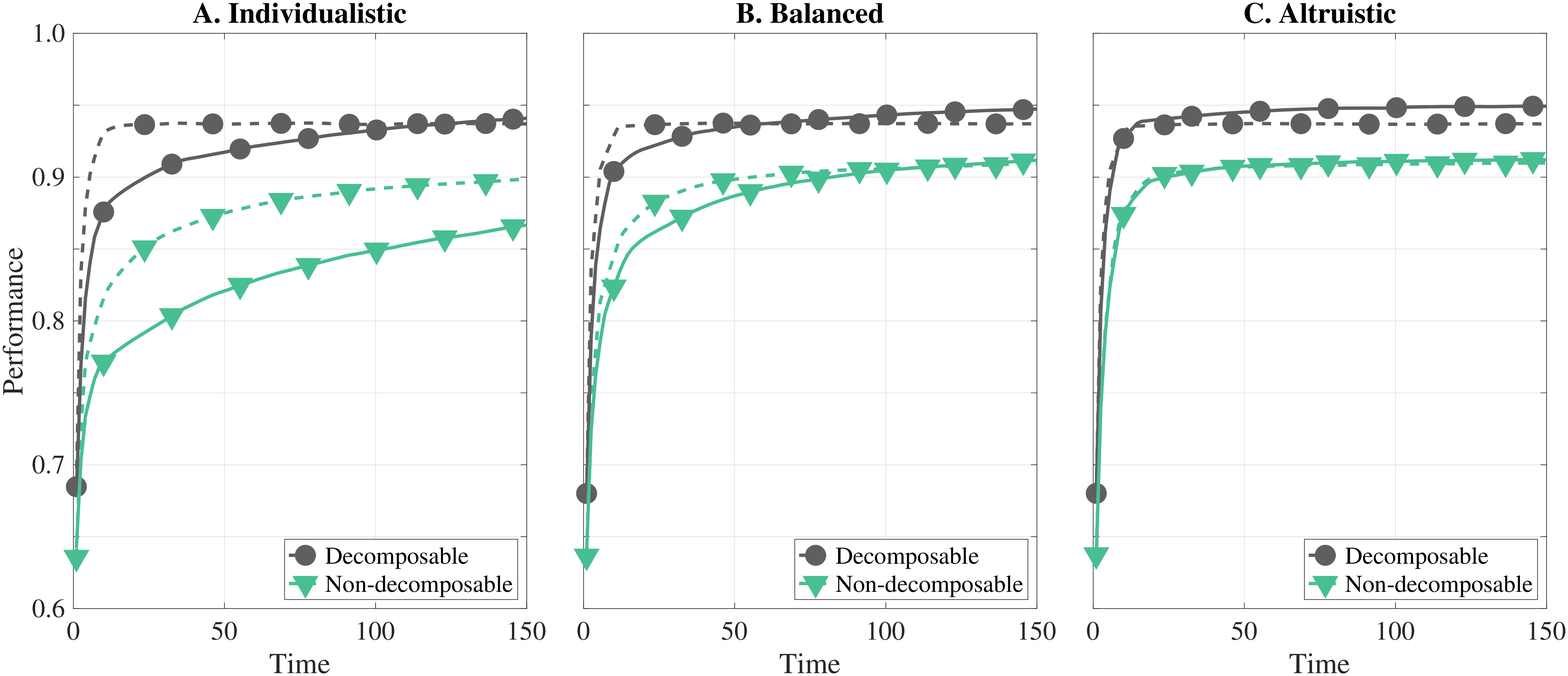}
\includegraphics[width=\textwidth]{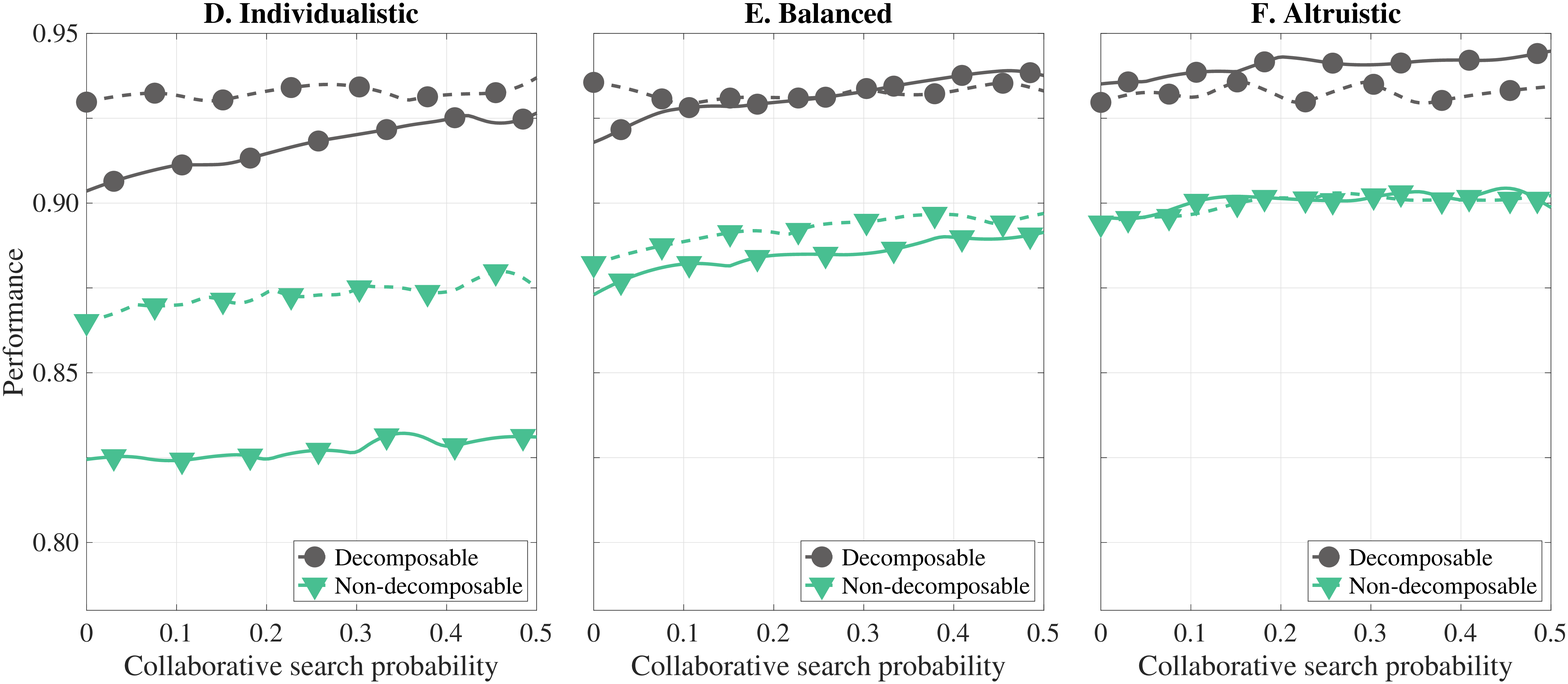}
\begin{scriptsize}Dashed ($--$) and solid lines (---) stand for benchmark scenarios and scenarios with task re-allocation, respectively. \end{scriptsize}
\caption{Partial dependence of performance on time and collaborative search probability}
\label{fig:results}
\end{figure}

The partial dependencies of performance on time and collaborative search probability are presented in Fig. \ref{fig:results}; top-down and bottom-up organizational designs are indicated by solid and dashed lines, respectively. Grey circles (\begingroup\color{gray}$\newmoon$\endgroup) indicate scenarios with decomposable tasks, and green triangles (\begingroup\color{green}$\blacktriangledown$\endgroup) stand for non-decomposable tasks.

For \textit{{decomposable tasks}}, the partial dependencies indicate that the performances in top-down designed organizations grow relatively fastly and reach the upper limit early in the observation period. For emergent organizational structures, both the speed and the upper limit are affected by the incentive parameter: The performance grows relatively slowly and eventually reaches the upper limit of the performance in top-down organizations if \textit{individualistic incentive systems} are effective (Fig. \ref{fig:results}.A). The partial dependencies of task performance on the collaborative search probability (Fig. \ref{fig:results}.D) indicate that this pattern is reinforced if the collaborative search probability is low, i.e., the distance between the performances in the two cases gets larger. In contrast, the performances become more similar if the collaborative search probability is high.  In the case of \textit{balanced incentive systems} (Fig. \ref{fig:results}.B), the dependence of the performance on time is relatively similar to panel A, and the collaborative search probability appears not to significantly affect the slopes of the performance curves (Fig. \ref{fig:results}.E). If \textit{altruistic incentive systems} are effective (Fig. \ref{fig:results}.C), the performance reacts more substantially to time when the organizational structure is dynamic. The performance is eventually higher compared to the performance in top-down designed organizations. The results presented in Fig. \ref{fig:results}.F indicate that this effect is reinforced if the collaborative search probability increases. This means that relatively high collaborative search probabilities pay off in performance if altruistic incentive schemes are effective in the organization.

The patterns observed for scenarios with {\textit{non-decomposable tasks}} are similar to those for decomposable tasks, whereby, as already evident from Fig. \ref{fig:results-complexity}, relatively lower performances are achieved. For \textit{individualistic incentive schemes} (Fig. \ref{fig:results}.A), the performance increases faster and reaches a higher level in top-down designed organizations than in cases with emergent structures; increasing the collaborative search probability in these cases only has negligible effects. The performance increases faster but has approximately the same upper limit if \textit{balanced and altruistic incentive mechanisms} are effective in the organization (Fig. \ref{fig:results}.B--C). When altruistic incentive systems are effective, the performances in top-down and bottom-up designed organizations become very similar; the partial dependencies plotted in Fig. \ref{fig:results}.F indicate that this pattern is robust against variations in the collaborative search probability.

\subsection{Task allocation efficiency}


This section analyses to what extent the emerging organizational structure in scenarios with task re-allocation conforms to the task allocation suggested by the mirroring hypothesis (i.e., the solid lines in Fig. \ref{fig:matrices}). The following task allocation efficiency in scenarios with top-down structures are used as a benchmark: In the case of decomposable decision problems, all interdependencies are internalized into the agents' decision problems (Fig. \ref{fig:matrices}, $K=2$), and, in consequence, the benchmark efficiency metric reaches a value of $1$. For non-decomposable decision problems, only a subset of the interdependencies can be internalized; only $6$ out of $15$ interdependencies ($40\%$) are inside an agents' decision problems in Fig. \ref{fig:matrices}, $K=5$, and, in consequence, the benchmark efficiency metric is $0.4$.

\begin{figure}
\centering
\includegraphics[width=0.95\textwidth]{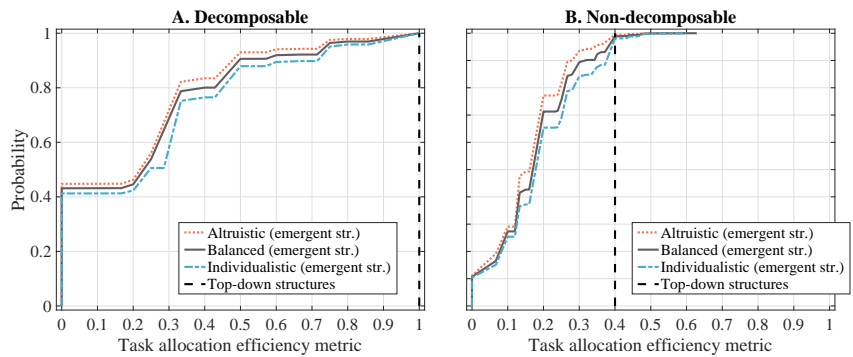}
    \caption{Cumulative distributions of the task allocation efficiency metric}
    \label{fig:ratios}
\end{figure}
The cumulative distributions of the task allocation efficiency metric are plotted in Fig. \ref{fig:ratios} (for all agents and all periods).
Interestingly, in only approx. $10\%$ of the cases, agents achieve a task allocation efficiency of $0.5$ out of $1$ in the case of decomposable tasks and $0.3$ out of $0.4$ for non-decomposable tasks. Even though the signals for task re-allocation are based on the agents' beliefs on interdependencies, the incentive parameter affects the task allocation efficiency: Irrespective of task complexity, altruistic incentive schemes result in a  slightly higher task allocation efficiency; this might be driven by an indirect effect coming from the individual search behavior induced by altruistic incentives as well as the resulting update of beliefs on interdependencies.

\section{Conclusions}
\label{sec:conclusion}

This paper presents a model of either dynamic or static organizations, in which search processes are carried out individually or collaboratively. The results indicate that collaborative search processes can indeed weaken the adverse effects of emergent task allocations that do not conform to the mirroring hypothesis. However, this is only true if there is a fit between the search processes and the remaining organizational design elements, namely with the inventive scheme: The results indicate that emergent approaches to organizational design work best with rather altruistic incentive schemes. Surprisingly, the results also indicate that organizations are better off if they follow an emergent design approach together with altruistic incentives if tasks are decomposable: In these cases, the performance even exceeds that of top-down organizations. Thus, the results indicate that the long standing finding that an organization's structure should mirror the technical interdependencies of the task the organization faces is not necessarily applicable in organizations with emergent structures.

This work can be seen as the first step toward an organizational design theory in dynamic organizations with autonomous agents. Further research could, for example, analyze different strategies for task re-allocation (e.g., different ways to compute the signals), different network structures for organizational links, and the effects of collaborative search in networks of organizations. Also, future research might take into account other forms of performance landscapes (e.g., plateaued landscapes).

%
%
%
\bibliography{bib}
\bibliographystyle{splncs03}

\end{document}